\documentclass{sig-alternate-05-2015}
\makeatletter
\def\ps@headings{%
\def\@oddhead{\mbox{}\scriptsize\rightmark \hfil \thepage}%
\def\@evenhead{\scriptsize\thepage \hfil \leftmark\mbox{}}%
\def\@oddfoot{}%
\def\@evenfoot{}}
\makeatother
\usepackage{algorithmic}
\usepackage{algorithm}
\usepackage{verbatim}

\usepackage{amsmath, amssymb}
\usepackage{enumerate}

\usepackage{eufrak}

\usepackage{verbatim}
\usepackage{array}
\usepackage{graphicx}
\usepackage{float}%\usepackage[tight,footnotesize]{subfigure}
\usepackage[footnotesize]{subfigure}

\usepackage{balance}

\newfont{\mycrnotice}{ptmr8t at 7pt}
\newfont{\myconfname}{ptmri8t at 7pt}
\let\crnotice\mycrnotice%
\let\confname\myconfname%

\begin{document}

\CopyrightYear{2016} 
\setcopyright{acmcopyright}
\conferenceinfo{MSWiM '16,}{November 13-17, 2016, Malta, Malta}
\isbn{978-1-4503-4502-6/16/11}\acmPrice{\$15.00}
\doi{http://dx.doi.org/10.1145/2988287.2989155}

\clubpenalty=10000 
\widowpenalty = 10000

\title{Making On-Demand Routing Efficient \\ with Route-Request Aggregation}

\author{
Maziar Mirzazad-Barijough
%$^1$
and J.J. Garcia-Luna-Aceves
\\
%$^{1,2}$ \\
%$^1$
Department of Computer Engineering,
 University of California, Santa Cruz, CA 95064\\
% $^2$Palo Alto Research Center, Palo Alto, CA 94304 \\
 Email: maziar@soe.ucsc.edu, jj@soe.ucsc.edu 
 }

\maketitle

\begin{abstract}
In theory, on-demand routing is very attractive for mobile ad hoc networks (MANET), because it induces signaling only for those destinations for which there is data traffic. However, in practice, the signaling overhead of existing on-demand routing protocols becomes excessive as the rate of topology changes increases due to mobility or other causes. We introduce the first on-demand routing approach that eliminates the main limitation of on-demand routing by aggregating route requests (RREQ) for the same destinations. The approach can be applied to any existing on-demand routing protocol, and we introduce the Ad-hoc Demand-Aggregated Routing with Adaptation (ADARA) as an example of how RREQ aggregation can be used. ADARA is compared to AODV and OLSR using discrete-event simulations, and the results show that aggregating RREQs can make on-demand routing more efficient than existing proactive or on-demand routing protocols.

\end{abstract}

\vspace{-0.1in}


\vspace{-0.08in}
%\category{C.2.1}{Network Architecture}{Network Architecture and Design}
%\category{C.2.2}{Network Protocols}{Routing protocols}

\vspace{-0.1in}
\terms{Theory, Design, Performance}

\vspace{0.1in}
 % NOT required for Proceedings

\section{Introduction}

Many routing protocols have been proposed for mobile ad-hoc networks (MANET), and
can be categorized as proactive, reactive, and hybrid routing protocols \cite{abolhasan, azzedine, prasant, adhoc-book, royer}. Proactive or table-driven routing protocols  maintain  routes  to every network destination independently of the data traffic being forwarded. Reactive or on-demand routing protocols maintain routes for only this destinations for which there are data packets to be forwarded. Hybrid protocols use proactive and on-demand mechanisms.  

Section \ref{MANET} provides a brief 
summary of the basic operation of proactive and on-demand routing. 
The proactive routing approach has the potential of high packet-delivery ratios and shorter end-to-end delays, because routes are established before data packets  requiring those routes are offered to the network. The price paid for such responsiveness is that signaling overhead is incurred even for those destinations that are not needed, which may be too high. 
In theory,  on-demand routing is designed to address this problem by requiring signaling overhead only for active destinations at the expense of incurring slightly longer latencies, because some data packets must wait for routes to be found.  However, as prior  comparative analysis of the performance of on-demand versus proactive routing schemes show \cite{Broch, clausen, jiang-comp, raju-comp, wu2010}, on-demand routing protocols end up incurring more overhead than proactive routing protocols  in MANETs when topology changes that impact existing data flows increase. 

Many techniques  (e.g., see \cite{abolhasan, azzedine, adhoc-book})
have been proposed to reduce the overhead incurred in the dissemination of each route request (RREQ), including  clustering, location information, dominating sets, and virtual coordinates. However, no prior work has addressed the impact of having relay routers aggregate RREQs they need to forward when they are intended for the same destinations.   

The main contribution of this paper is the introduction of a fault-tolerant approach for 
routers to aggregate RREQs originated by different sources and intended for the same destinations. The proposed {\em route-request aggregation} approach can be applied to any on-demand routing protocol
(e.g., AODV or DSR \cite{adhoc-book}) and can make any routing protocol that uses  on-demand routing techniques more efficient.

Section \ref{ADARA} introduces the Ad-hoc Demand-Aggregated Routing with Adaptation (ADARA) protocol as a specific example of the RREQ aggregation approach. 
Like AODV, ADARA uses destination-based sequence numbers to prevent routing-table loops  and request identifiers to denote each RREQ uniquely as in AODV. ADARA  introduces route-request aggregation and the use of broadcast signaling packets (RREQs, route replies and route errors) to substantially reduce signaling overhead.

Section  \ref{example} presents an example of the operation of ADARA and  how it improves performance compared to AODV \cite{adhoc-book}. However, the approach used in ADARA can be applied with proper modifications to on-demand routing based on source routes (e.g., DSR \cite{adhoc-book}) or path information \cite{olive}. It can also  be used in combination with prior techniques aimed at reducing signaling overhead, such as the use of  geographical coordinates of destinations \cite{lar, wang-13}, virtual coordinates, connected dominating sets \cite{wu99}, address aggregation \cite{Shiflet}, and clustering \cite{azzedine, hsr}.

Section \ref{performance} presents the results of simulation experiments used to compare ADARA with two routing protocols that are representative of the
state of the art in proactive routing and on-demand routing for MANETs, namely OLSR \cite{olsr} and AODV. The experiments were designed to study  the impact of node speed, pause times, number of sources, and network size on the  packet-delivery ratio, average end-to-end delay, and  signaling overhead. 
The results show that ADARA performs   better than OLSR and AODV in all cases. The key reason for this is that ADARA is able to establish routes on demand incurring far less overhead than AODV and OLSR.

\section{Related Work}
\label{MANET}

Many MANET routing protocols have been proposed since the introduction of the routing protocol for the DARPA packet-radio network \cite{jubin} and excellent surveys and comparative studies of this prior work  have been presented over the years \cite{abolhasan, azzedine, jiang-comp, prasant, adhoc-book, raju-comp, royer, wu2010}. 

OLSR is the best-known example of  proactive routing for MANETs \cite{olsr}. It uses HELLO messages to maintain neighbor connectivity, and Topology Control (TC) messages to disseminate  link-state information throughout the network. To reduce signaling overhead, OLSR takes advantage of connected dominating sets. Some  nodes are elected as multipoint  relays (MPRs) and only MPRs forward TC messages, and  only link-state information needed to connect MPRs is advertised in the network.

AODV \cite{adhoc-book} is the most popular example of the on-demand routing approach. To find a route to an intended destination,  a source broadcasts a RREQ stating the source and destination nodes, the most recent  sequence number known for each, a
a broadcast ID, and a hop count to the source. A router that forwards a RREQ for the first time 
creates a record for the RREQ  stating the source and broadcast-ID pair of the RREQ; and a
a reverse route to the source of the RREQ stating the 
next hop and hop count to the source, and the sequence number of the source. It maintains any RREQ record and reverse route for a finite time.
A router discards any  received RREQ that states a source and broadcast-ID pair for which it has a RREQ record. 

The intended destination or a router with a valid route to the destination responds to the RREQ by  sending a route reply (RREP) over the reverse route from which the RREQ was received. The RREP states the destination and the source of the RREQ, the destination sequence number, and the hop count to the destination. A router receiving a RREP establishes a route record to the destination stating the destination sequence number, the next hop to it, and the neighbors using the route (precursors). A router forwards only the first copy of a RREP 
(based on the destination sequence number) and increments by one the hop count to the destination when it forwards a RREP.

Link failures  can be recognized in AODV by the absence of HELLO messages sent periodically between neighbors.
When a node detects a link failure, it sends a route error (RERR)  to all neighbor nodes 
that are precursors of a route that is broken because of the link failure. Nodes receiving a RERR message invalidate all routes that were using the failed link and propagate the RERR message to their precursor nodes. 

Hybrid routing protocols attempt to reduce the signaling overhead of proactive and on-demand schemes by combining the two. This has been done by either using clusters within which routes to destinations are maintained proactively and using on-demand routing across clusters (e.g., ZRP \cite{adhoc-book}), or by maintaining routes to certain destinations proactively and using on-demand routing for the rest of destinations \cite{roy}.  

Interestingly, all prior approaches proposed  for on-demand and hybrid routing have assumed that a router that receives route-requests (RREQ) regarding destinations for which it does not have valid routes forwards {\em each} new RREQ  it receives, and replicas of the same RREQ are silently dropped.  This constitutes a major performance limitation for all on-demand and hybrid routing schemes proposed to date. Intuitively, as the number of destinations increase, the failure of just a few links may cause many sources to engage in the discovery of new routes to those destinations, with each source flooding RREQs.
Because a router forwards each RREQ it receives as long as it does not state the same source and request ID pair, the flooding of RREQs grows linearly with the number of sources,  even if the sources are seeking the same few destinations. 

The following section describes our approach to address this problem crated by too many RREQs. We use a specific protocol as an example of the basic approach.

% \vspace{-0.1in}
\section{ADARA}
\label{ADARA}

The design rational for ADARA is twofold. First, for the performance of an on-demand routing protocol to be comparable to or better than the performance of a proactive routing protocol,
the number of RREQs that sources 
initiate  in the route-discovery process must be kept to a minimum
when the network supports many data flows  and experiences topology changes.
Second, if the number of data flows intended for the same destination node is larger than the number of neighbors of that destination, the routes from the sources of the flows to the destinations  {\em must} have some routing relays in common. Accordingly, allowing routers to aggregate RREQs  intended for the same destination is bound to have a positive effect on the overall performance of the network.

ADARA (Ad-hoc Demand-Aggregated Routing with Adaptation) is the first on-demand routing protocol in which a router  aggregates RREQs from different sources intended for the same destination. 

ADARA adopts the use  of destination-based sequence numbers  as  in AODV to avoid routing-table loops, as well as the use of the source address and a request identifier created by the source to identify each RREQ. Other approaches have been proposed to avoid routing-table loops when routers maintain routes  on-demand \cite{stop, olive, mosko, rang05, roam} and can be used instead of the specific approach based on destination sequence numbers assumed in this paper. 

\subsection{Information Exchanged and Stored}

ADARA uses four types of signaling packets, all of which are sent in broadcast mode.

A Route Request ({\bf RREQ})  is denoted by $REQ[RID, $ $o,$ $ on,$ $ d, dn, $ $ho, $ $HSN ]$  and 
contains: 
A request identifier ($RID$), the address of the origin or source of  the RREQ  ($o$), 
a sequence number created by the origin ($on$), 
the address of the intended destination ($d$), the most recent sequence number known from $d$ ($dn$), a hop count to the origin of the RREQ ($ho$), and a HELLO sequence number $(HSN$).

A Route Reply ({\bf RREP}) is denoted by $REP[ d$ $, dn,$ $ hd, $ $LDN, $ $HSN ]$  and 
contains:  the address of the destination ($d$), the most recent sequence number known from $d$ ($dn$),  a hop count to the destination ($hd$), 
a list of designated neighbors ($LDN$) from which valid RREQs for  destination $d$ have been received, and a HELLO sequence number $(HSN$).

A Route Error ({\bf RERR}) is denoted by $RE[ HSN, LUA ]$  and
contains a HELLO sequence number $(HSN$) and a list of unreachable addresses ($LUA$).

A Hello message ({\bf HELLO})  is denoted by $H[HSN]$ and contains the sequence number of the sending node.

Each  router $i$ maintains a routing table ($RT^i$) and a pending request table ($PRT^i$). 
Each entry of $RT^i$ specifies: the address of the destination, a sequence number created by the destination,  a hop count to the destination, next hop to the destination, a list of precursor neighbors for the destination, and a lifetime.

$PRT^i$ is used to keep track of the RREQs received by router $i$, aggregate RREQs received for the same destination, and discard duplicates of the same RREQ. 
An entry in $PRT^i$ lists a destination address,  a list of precursor tuples, and a lifetime.
Each precursor tuple  consists of: the address of an origin node, the RID stated by that node, and the address of the precursor neighbor from which a RREQ was received.

ADARA is a soft-state protocol. Each entry in $PRT^i$ and $RT^i$ has a finite lifetime to allow router $i$ to
delete entries that become obsolete as a result of topology changes (e.g., the network is partitioned or a node fails).

\subsection{Updating Neighbor Connectivity}

Whenever a router receives a Hello message, a RREQ, a RREP, or  a RERR, it calls 
the Hello Process function shown in Algorithm \ref{Process Hello}
to update routes to neighbor routers.  This process uses the HSN
included in each signaling packet. The HSN a router includes in a RREQ, RREP or RERR is simply the value of its current sequence number.

{\fontsize{8}{8}\selectfont
\begin{algorithm}[h]
\caption{Processing Hello}
\label{Process Hello}
{\fontsize{8}{8}\selectfont
\begin{algorithmic}
\STATE{{\bf function}  Process\_Hello}
\STATE {\textbf{INPUT:}  $sender$, $r\_table^i$, $HelloSeqNo$;}

\STATE{$route = r\_table^i.lookup(sender);$}
\STATE{$route.setHop(1);$}
\STATE{$route.SetDes(sender);$}
\STATE{$route.SetNextHop(sender);$}
\STATE{$route.SetSeqNum(HelloSeqNo);$}
\STATE{$route.mark(Valid);$}
\STATE{$r\_table.update(route);$}

\end{algorithmic}
}
\end{algorithm}
}

\subsection{Route Discovery Process}

A router originates a RREQ when it has no valid route to an intended destination as a result of topology changes or because a new destination is of interest to the router.
Algorithm  \ref{Process RREQ} shows the steps taken by a router to process a RREQ it receives from a neighbor.

After the neighbor information is updated according to  Algorithm \ref{Process Hello},
router $i$ updates its routing information regarding the origin of the RREQ. 
Router $i$ uses Algorithm \ref{Aggregate RREQ} to process the RREQ based on its origin,
the RID created by the origin, and the entries in $PRT^i$.  

Router $i$ sends back a RREP to the RREQ it receives if it is the intended destination or
$RT^i$ contains a valid entry for the destination stated in the RREQ with a sequence number  that is higher than or equal to the  destination sequence number stated in the RREQ. The RREP is broadcast to all neighbors
and states the hop count to the destination, the destination sequence number, a HELLO sequence number for itself, and the list of designated neighbors. 

If router $i$ has no valid route to the intended destination in the RREQ and  there is no entry in $PRT^i$ for  that destination, router $i$ 
creates a $PRT^i$ for the destination and broadcasts the RREQ to its neighbor routers 
with its own HSN and its own hop count to the origin of the RREQ.
On the other hand, if there is an entry for the destination in $PRT^i$, there are various cases to consider. 

If the RREQ  is a replica of a RREQ received from the same origin
(i.e., there is a pending RREQ for the destination from the same origin and with the same RID),  the RREQ is silently dropped.
If the RREQ is not a replica of a RREQ already received, but is a retransmission of a RREQ from one of the origins of the request, it means that the origin is retransmitting its RREQ due to a timeout expiration. 
Accordingly, router $i$  updates the RID of the corresponding precursor tuple and broadcasts the RREQ to its neighbor routers.
Lastly, if the  RREQ is from a different source than those listed in $PRT^i$, router $i$ simply adds a precursor tuple $PRT^i$ with the address of the origin, the RID created by the origin, and the address of the neighbor that sent the RREQ. We say that the   RREQ is aggregated in such a case.

{\fontsize{8}{8}\selectfont
\begin{algorithm}[h]
\caption{Process RREQ  from router $s$ at router $i$}
\label{Process RREQ}
{\fontsize{8}{8}\selectfont
\begin{algorithmic}
\STATE{{\bf function}  Process\_RREQ}
\STATE {\textbf{INPUT:}  $rreq$,$org$, $r\_table^i$,$Destination$;}
\STATE {$des=rreq.getDestination();$}
\STATE {$processHello (s, RREQ.HelloSeqNo);$}
\STATE{$aggregated=PRT.Aggregate(RREQ);$}
\STATE{$UpdateReversePath(RREQ, org);$}
\STATE{$rt=r\_table^i.lookup(des);$}
\IF{$(rt) \land (rt.seq \geq rreq.Seq) \land (rt == VALID)$}
	\STATE{$rrep=create\_rrep(rt);$}
	\STATE{$rrep.SetHelloSeq(LocalSeq)$}
	\STATE{$Broadcast(rrep);$}
\ELSE
	\IF{$!aggregated$}
		\STATE{$rreq.SetHelloSeq(LocalSeq)$}
		\STATE{$Broadcast(rreq);$}
	\ENDIF
\ENDIF

\end{algorithmic}
}
\end{algorithm}
}

\vspace{-0.2in}
{\fontsize{8}{8}\selectfont
\begin{algorithm}[h]
\caption{Aggregate RREQ  $i$}
\label{Aggregate RREQ}
{\fontsize{8}{8}\selectfont
\begin{algorithmic}
\STATE{{\bf function} Aggregate\_RREQ}
\STATE {\textbf{INPUT:}  $rreq$, $PRT^i$;}
\STATE {$des=rreq.getDestination();$}
\STATE {$org=rreq.getOrigin();$}
\STATE {$id=rreq.getId();$}
\IF{$\exists enrty \in PRT \land entry_{org}=org \land entry_{id}=id $}
	\STATE{$drop(rreq);$} //Duplicate RREQ
	\STATE{$return$ $true;$}
\ENDIF
\IF{$\exists e \in PRT \land e_{org}=org \land e_{des}=des \land e_{id} \neq id$}
		\STATE{$update(entry,rreq);$} //Retransmitted RREQ
	 	\STATE{$return$ $false;$}
\ENDIF
\IF{$\exists e \in PRT \land e_{org} \neq org \land e_{des}=des$}
		\STATE{$PRT.AddEntry(rreq);$} // Aggregate
	 	\STATE{$return$ $true;$}
\ENDIF
\STATE{$PRT.AddEntry(rreq);$}
\STATE{$return$ $false;$}

\end{algorithmic}
}
\end{algorithm}
}

When router $i$ receives a RREP, it updates its neighbor information according to Algorithm \ref{Process Hello}. Router $i$ accepts the information in the RREP and updates $RT^i$ for the destination stated in the RREP
if either the destination sequence  number  is higher than the destination sequence number in $RT^i$ or the sequence numbers are the same but the hop count to the destination in the RREP is smaller than the 
corresponding hop count in $RT^i$.  

 For the case of a valid RREP, router $i$ creates or updates the entry in $RT^i$  for the destination. The entry states  the destination sequence number obtained in the RREP, its hop count to the destination, and the list of precursor neighbors for the destination.
The  precursor neighbors are simply those neighbors listed in precursor tuples for the destination in $PRT^i$.  If the router is a member of LDN of RREP, then the router $i$  broadcasts the RREP to its neighbors stating its own hop count to the destination, its own HELLO sequence number, and a list of designated neighbors of  router $i$ that need to process and perhaps forward the RREP.
Router $i$ can then delete the entry for the destination in $PRT^i$ . In case the router is not in LDN, after updating the routes, router will drop the RREP to limit the region within which the RREP is re-broadcast.

{\fontsize{8}{8}\selectfont
\begin{algorithm}[h]
\caption{Processing RREP  from router $s$ at router $i$}
\label{Process RREP}
{\fontsize{7}{7}\selectfont
\begin{algorithmic}
\STATE{{\bf function}  Process\_RREP}
\STATE {\textbf{INPUT:}  $rrep$,$sender$, $r\_table^i$,$Destination$;}
\STATE {$des=rrep.getDestination();$}
\STATE {$processHello (sender, rrep.GetHelloSeqNo());$}
\STATE{$rt=r\_table^i.lookup(des); $}
\STATE{$intended=false;$}
\IF{$currentNode \in RREP.LDN()$}
	\STATE{$designated=true;$}	
\ENDIF
\IF{($rt\_des \neq empty$)}
	\IF{$ (rrep.seq > rt\_des.seq) \lor  (rrep.seq=rt\_des.seq\land rrep.hop<rt\_des.hop)$} 
		\STATE{$rt\_des.update(rrep);$}
	\ENDIF
\ELSE 
		\STATE{$rt\_des = r\_table^i.AddRoute(RREP);$}
\ENDIF
\IF{$designated \neq true \land PRT.lookup(Des_{RREP}).Count \leq 1$}
	\STATE{$return;$}
\ENDIF

\STATE{$RREP.ClearLDN();$}
\FOR{$\textbf{each}$ $entry \in PRT.lookup(Des_{RREP})$}
	\STATE{$PRT.remove (des);$}
	\STATE{$rt\_org=r\_table^i.lookup(entry.org);$}
	\STATE{$RREP.LDN.Add (entry.PrecursorNeighbor);$}
	\STATE{$rt\_des.AddPrecursor(entry.PrecursorNeighbor)$}
\ENDFOR
\STATE{$RREP.setHelloSeq(LocalSeq);$}
\STATE{$Broadcast(RREP);$}

\end{algorithmic}
}
\end{algorithm}
}

\vspace{-0.2in}
{\fontsize{8}{8}\selectfont
\begin{algorithm}[h]
\caption{Processing RERR  from router $s$ at router $i$}
\label{Process RERR}
{\fontsize{8}{8}\selectfont
\begin{algorithmic}
\STATE{{\bf function}  Process\_RERR}
\STATE {\textbf{INPUT:}  $rerr$, $r\_table^i$,$unreachable$;}
\STATE {$processHello (s);$}
\STATE {$rtList = $ Get All entries in $r\_table^i$ that use $s$ toward unreachable routers;}
\STATE{$hasPrecursor=false;$}
\FOR{$\textbf{each}$ $rt \in rtList$}
	\IF{$rt.precursorCount()>0$}
		\STATE{$hasPrecursor=true;$}
	\ENDIF
	\STATE{$invalidate(rt);$}
\ENDFOR
\IF{$hasPrecursor$}
	\STATE{$RERR.SetHelloSeq(LocalSeq);$}
	\STATE{$Broadcast(RERR);$}
\ENDIF

\end{algorithmic}
}
\end{algorithm}
}

\subsection{Handling Errors and Topology Changes}

Route error messages are created when no route is found toward a destination router or a link break is detected. A router assumes that a link with a neighbor  is down when it fails to receive any signaling packet within interval defined for the reception of signaling packets from a neighbor. An error message states all the destinations for which routes are broken as a result of the link failure.

Algorithm \ref{Process RERR} shows the steps taken by router $i$ to process a RERR from a neighbor. Router $i$ invalidates all the routes to destinations listed in the RERR that require the router sending the RERR as the next hop. 
Router $i$   broadcasts  a RERR  it receives if at least one precursor neighbor exists for
the destinations listed in the RERR.  Accordingly, only routers that established routes to destinations by forwarding RREQs may have to forward RERRs.

\section{ADARA Example}
\label{example}

\begin{figure}[h]
  \includegraphics[scale=0.35]{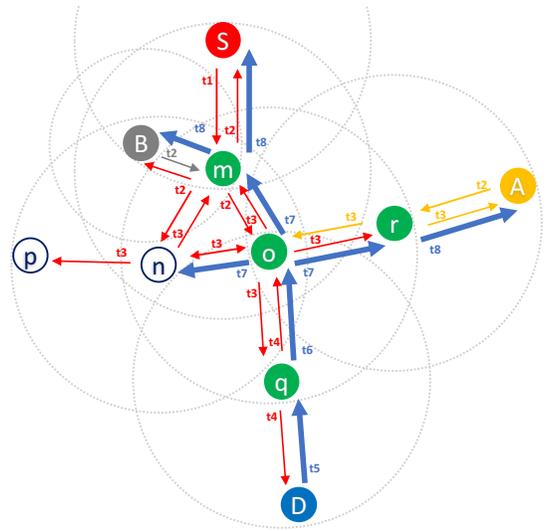}
  \caption{Dissemination of RREQs and RREPs in ADARA}
  \label{topology}
\end{figure}

Figure \ref{topology} shows a small wireless network in which ADARA is used. The network consists of six relay routers ($m$, $n$, $o$, $p$, $q$, and $r$), three source routers ($S$, $A$, and $B$), and one destination router $D$. The example assumes that no router has valid routes for destination $D$, and shows router $S$ generating and broadcasting a RREQ for destination $D$ at  time $t_1$.
The propagation of this RREQ is indicated by thin arrows in the figure, and the propagation of RREPs is shown with thick blue arrows.
The RREQ from router $S$ states 
$REQ[RID_S, $ $S,$ $ on_S,$ $ D,$ $ dn = 0, $ $ho = \infty, $ $HSN_S ]$

When router $m$ receives the RREQ from source $S$, it adds a route for destination router $S$ as a destination in $RT^m$ with a hop count of one and  $S$ as the next hop. Router $m$ also  creates an entry for $D$ in  $PRT^m$  listing the precursor tuple $[S, RID_S, S]$, which  states $S$ as the origin of the RREQ with a RID equal to $RID_S$ and source $S$ as the neighbor from which the RREQ was received.

The example shows routers $A$ and $B$ originating  RREQs for destination $D$ at time $t_2 > t_1$. As the figure shows, router $r$ forwards the RREQ at  time $t_3 > t_2$.
However, when router $m$ receives the RREQ from router $B$ for  destination $D$ shortly after time  $t_2$, it simply aggregates the RREQ, because $PRT^m$ contains an entry for $D$.   Router $m$ does this by adding the precursor tuple 
$[B, RID_B, B]$ to its entry for destination $D$ in  $PRT^m$.

Router $o$  creates an entry for $D$ in $PRT^o$
after receiving the RREQ  forwarded by router $m$, and that entry lists the precursor tuple $[S, RID_S, m]$. Accordingly, when router $o$ receives the RREQ forwarded by router $r$ shortly after time $t_3$, it can simply aggregate the RREQ.  It does this by adding the precursor tuple $[A, RID_A, r]$ to the entry for destination $D$ in  $PRT^o$. 
Similarly, when router $r$ receives the RREQ forwarded by router $o$ (originated by source $S$) shortly after time $t_3$, it already has an entry for destination $D$ in $PRT^r$ and hence aggregates the RREQ received from router $o$ by adding the precursor tuple $[S, RID_S, o]$ to its list of precursor tuples for destination $D$.

We note that, shortly after time $t_3$, routers $n$ and $o$ receive the RREQ originated by source $S$ from each other. Both routers simply ignore the replicas of the RREQ originated by  router $S$ because they each have an entry for destination $D$ in their PRTs listing a precursor tuple with the same source router and source sequence number than the ones included in the RREQ they receive from each other. 

As the RREQs from sources $S$, $A$, and $B$ are disseminated in the network, relaying routers add precursor tuples to their PRTs for destination $D$. These tuples allow each relay router to decide whether to broadcast a RREP for $D$ when it receives a RREP from a neighbor. Destination $D$ generates a RREP for itself at time $t_5$ when it receives the RREQ from router $q$.  Starting with router $q$, the RREP is disseminated back to the sources that originated RREQs for $D$ along the reverse paths traversed by the RREQs thanks to the precursor tuples maintained in the PRTs of routers. Each relaying router re-broadcasts the RREP for destination $D$  if it has at least one precursor tuple for $D$ in its PRT, which results in RREPs being disseminated along a directed acyclic graph as illustrated in Figure 1. Each router that forwards a RREP copies the  precursor neighbors for $D$ to its RT. 

A RREP contains the list of designated neighbors (LDN) that may  forward the RREP as needed, and is based on the precursors stated for a given destination in the PRTs of routers. In the example, the LDN of the RREP from router $q$ lists router $o$, and the LDN of the RREP from router  $o$ states routers $m$ and $r$.  
Accordingly, as shown in Figure 1, when router $n$ receives the RREP from router $o$, it does not forward the RREP, given that it is not listed in the LDN of the RREP from router $o$. However, router $n$ adds a routing entry for $D$ in $RT^n$. 
Routers $m$ and $r$ forward the RREPs they receive from router $o$.

Router $r$ forwards the RREP with an LDN listing routers $o$ and $A$. Router $o$  simply ignores the RREP from $r$, and source $A$ is able to start sending data packets to $D$. By the same token, routers that receive RREPs from the next hops to the sources of RREQs can ignore the RREPs because they are not listed in the LDNs of those RREPs. 

In contrast to the above, AODV and other on-demand routing protocols would require the dissemination of the RREQs from $S$, $A$ and $B$ throughout the  entire network, and for each origin router, a RREP would be sent on the path from source to destination. 

\begin{figure}[h]
   \includegraphics[scale=0.4]{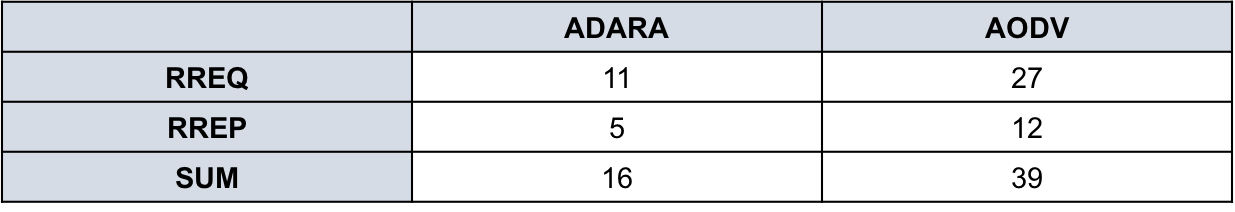}
   \caption{Number of Signaling Packets Sent for ADARA and AODV}
 \label{compare}
\end{figure}

 Figure \ref{compare} shows the number of signaling packets sent in the topology of Figure \ref{topology}. The number of RREQs in ADARA is much smaller  compared to AODV, which is a direct result of RREQ aggregation. Using ADARA, the RREQ generated by $A$  is only sent by routers $A$ and $r$, and the RREQ from router $B$ is sent once by router $B$ and aggregated at router $m$. On the other hand, using AODV, the RREQs from routers $S$, $B$, and $A$ are   flooded in the network. The number of RREPs sent over the network in ADARA is also lower than AODV as a result of the   aggregation or RREQs. In ADARA, RREPs are sent once on the path up to an aggregation point. In AODV, each RREP sent once on each path. As a result, the number of RREQs and RREPs in AODV is 2.5 times larger than in ADARA for this example.  Furthermore, since all RREPs are broadcast messages, routers on the path from a source to a destination do not generate Hello messages for a time interval. For the case of AODV, Hello messages are generated independently of the RREPs being  sent. 

 \begin{figure*}
  \includegraphics[width=\textwidth,height=4cm]{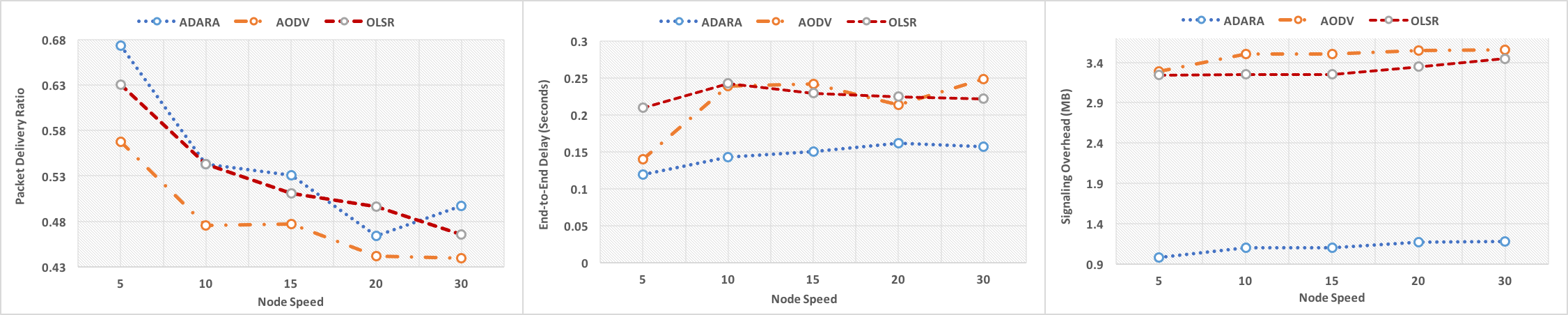}
  \caption{Performance comparison as  a function of router speed.}
  \label{sim speed}
\end{figure*}

%\vspace{-0.1in}
\begin{figure*}
  \includegraphics[width=\textwidth,height=4cm]{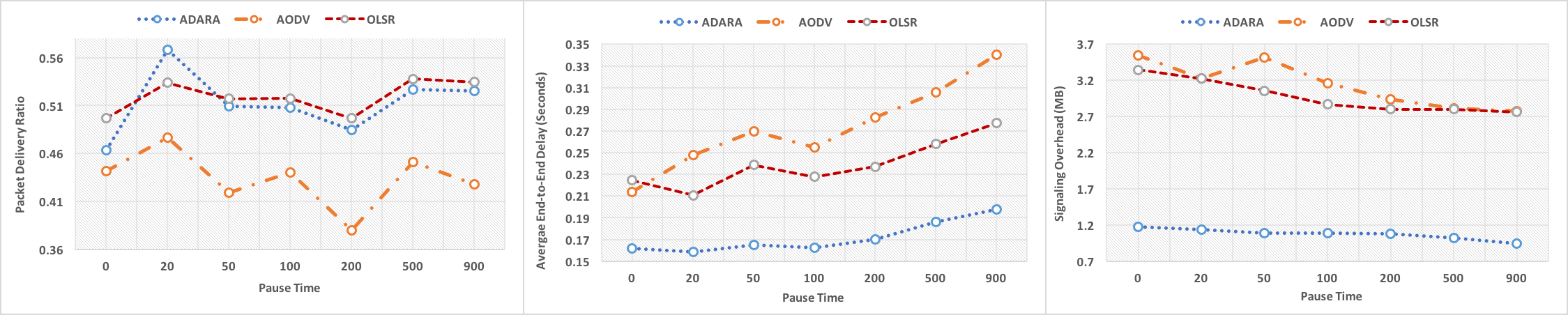}
  \caption{Performance comparison as  a function of Pause Time.}
  \label{sim pause}
\end{figure*}

%\vspace{-0.1in}
\begin{figure*}
  \includegraphics[width=\textwidth,height=4cm]{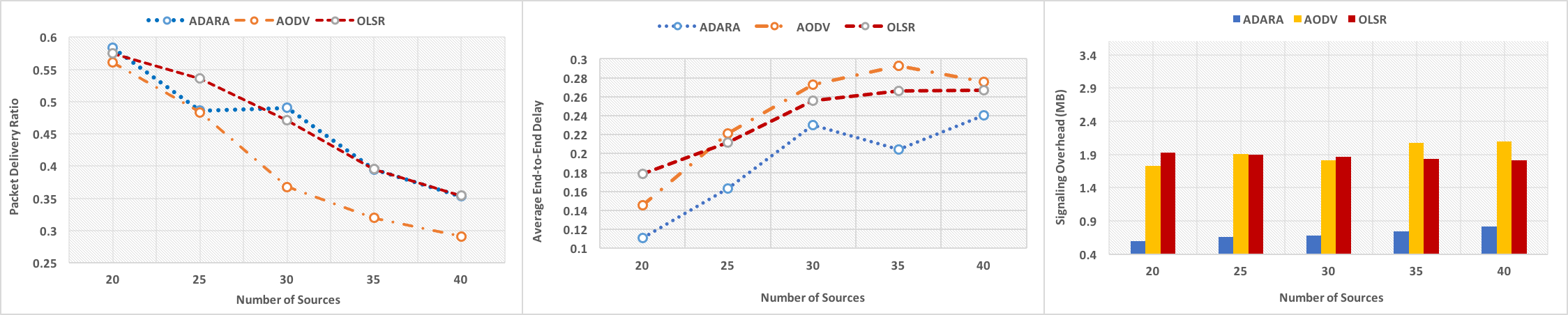}
  \caption{Performance comparison as  a function Number of Sources.}
  \label{sim flow}
\end{figure*}

%\vspace{-0.1in}
\begin{figure*}
  \includegraphics[width=\textwidth,height=4cm]{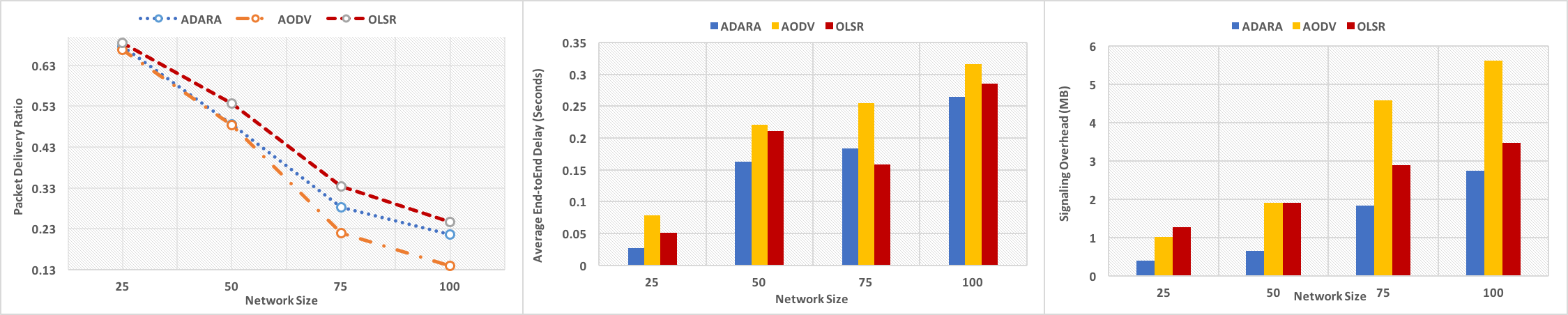}
  \caption{Performance comparison as  a function Network Size.}
  \label{sim networkSize}
\end{figure*}

\section{Performance Comparison}
\label{performance}

\subsection{Simulation Model and Parameters}

\begin{figure}[h]
  \includegraphics[scale=0.43]
  {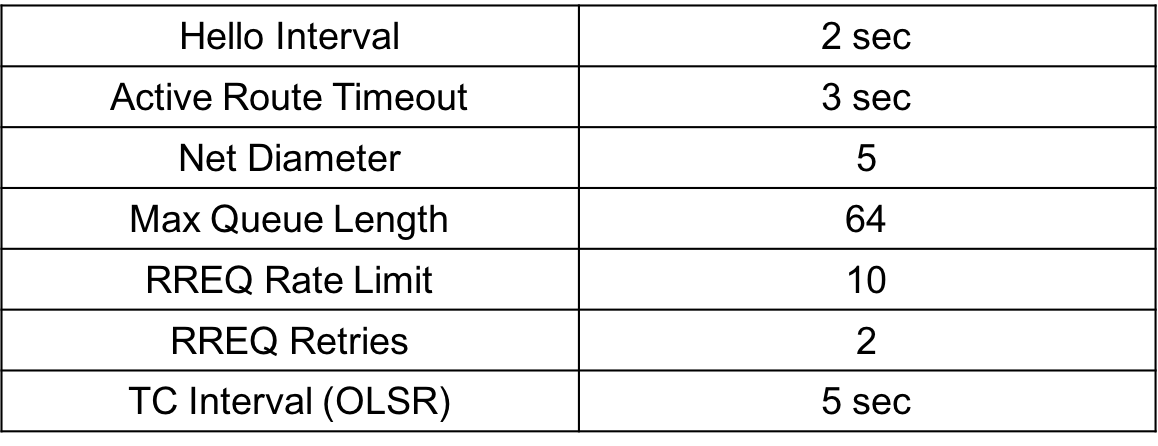}
  \caption{Simulation Configuration for ADARA, AODV, and OLSR.}
  \label{SimSet}
\end{figure}

We implemented ADARA in ns3 and used the ns3 implementations of AODV and OLSR without modifications to compare their  performance. 
Figure \ref{SimSet} shows  simulation-environment settings for AODV, OLSR, and ADARA.

The Distributed Coordination Function (DCF) of IEEE 802.11n  2.4 Ghz with rate of 2Mbps is used as the MAC-layer protocol for unicast data transmission.   To avoid incorrect paths resulting from transmission-range differences between unicast and broadcast transmissions, we made sure that both broadcast and unicast packets are sent with the same rate (2Mbps) and range.  Transmission power is adjusted to fix the transmission range to 250 meters.
Both AODV and ADARA use a sending buffer of 64 packets. These buffers  store packets waiting for RREP message to the desired destination for 30 seconds. 

Simulations include 50 routers spread uniformly  in a $300m \times 1500m$ area. For other scenarios, 25, 75, and 100 routers are uniformly spread in a $300m \times 1500m$, $300m \times 1500m$, and $300m \times 1500m$ respectively. Routers use the random-waypoint mobility model with a randomly-chosen moving speed between 0 and 20 m/s and pause time of 0 seconds. A router chooses a destination location randomly and moves toward that destination with a randomly chosen speed between zero and the specified maximum speed. When a destination location is reached, the router  remains there for a specified pause time.   

The scenarios include 25 data flows from 25 different source routers. The destination router for each flow is a specific router with probability 0.5 and is chosen randomly from all routers with probability 0.5. 
Traffic sources are on-off applications with on and off time of 1 second, which generate packets of size 512 bytes and rate of 15 packets per second.
For network sizes of 25 and 50 routers, simulations are run for 900 seconds, and for networks sizes of 75 and 100 routers, the simulation time is 500 seconds. 

The signaling overhead in AODV includes its five types of packets:  RREQ, RREP, RERR, Route Reply Ack, and HELLO messages. In OLSR, the signaling overhead includes, Topology Control (TC) messages and HELLO messages. In  ADARA, the signaling overhead includes all its different types of signaling packets.

We compared ADARA, AODV and OLSR  based on 
the  packet delivery ratios (PDR), the average end-to-end delay, and the number of signaling packets  sent by all routers. PDR indicates the number of packets received by destination routers divided by number of packets sent by the source routers. The average end-to-end delay is the time elapsed from the time a packet is sent by a source until it is received by its destination. For the case of ADARA and AODV, this delay includes the duration packet is buffered waiting for RREPs.  

The scenarios used to compare the three routing protocols were chosen to stress all three protocols, rather than to attain good performance for either on-demand or proactive routing.

\subsection{Effect of Mobility}

In this scenario, 50 routers are spread randomly in a $300m \times 1500m$ area, with  25 of the routers generating traffic, each with 15 packets per second. The destination  for each flow is a specific router with probability of 0.5 and is chosen randomly from all routers with probability of 0.5.  We tried different maximum mobility speeds of 5 m/s to 30 m/s with a zero pause. 

Figure \ref{sim speed} shows the PDR,  average end-to-end delays, and the signaling overhead incurred by the three protocols as a function of router speed. 

Higher router speed results in  more topology changes. 
The drastic drop in PDR in all protocols  is due to routes breaking due to router mobility and the time needed by the routing protocols to obtain new routes.
OLSR requires  routers to detect link failures and additions based on the absence or reception of  HELLO messages, and TC messages to inform all routers of the topology  so that new routes can be established. 
Given that  TC messages are sent periodically listing one or multiple link states, the signaling overhead in OLSR remains fairly constant as a function of router speed. However, this means that more link changes  take place between periodic transmissions of TC messages as router speed increases, which results in more data packets being lost as they traverse paths that are broken. 

Link failures in AODV and ADARA are detected by the absence of  a number of consecutive Hello messages, and a route discovery process is done to inform sources of new routes to destinations. Because of the delays incurred in detecting link failures and in establishing new routes after that, as router speed increases more and more data packets traversing failed routes end up being  dropped.  

The lower delays obtained with ADARA can be attributed to the aggregation of RREQs, which reduces the number of RREQs being flooded and hence reduces network congestion, as well as the fact that each signaling packet is sent in broadcast mode containing the current sequence number of the transmitting router, which helps routers detect link failures and repair routes more quickly.

The enormous impact of RREQ aggregation in ADARA is evident in Figure \ref{sim speed}.
In OLSR, TC messages must be disseminated by MPRs throughout the network and in AODV, each RREQ is  flooded throughout the network. By contrast, a RREQ in ADARA is disseminated throughout the network only when no other RREQ asking for a route for the same destination has been forwarded recently.  The size of TC messages in OLSR is much larger than the size of RREQs in AODV, which accounts for the similarity in signaling overhead between the two even though routers in OLSR disseminate fewer signaling packets than in AODV.

Figure \ref{sim pause} shows the performance comparison of the three protocols as a function of mobility pause time. For this simulation runs, we considered 50 routers in a area of $300m \times 1500m$ with 25 flows as described previously.  Pause times vary from 0 seconds (i.e., constant mobility) to to 900 seconds (i.e., almost static routers).  The speed of routers  is chosen randomly between zero and  20 meters per second. 
As it is can be seen from the figure, the packet delivery ratio for ADARA is very close to that of OLSR, while AODV is much lower for all pause times.  
Average delays in ADARA are much lower than those attained in OLSR and AODV over all pause times, and AODV renders the higher delays in all cases.
It is also evident that ADARA incurs far less signaling overhead than OLSR and AODV for all pause times. As should be expected, less signaling overhead is incurred by all protocols as the pause time increases.  

\subsection{Effect of Number of Flows}

Figure \ref{sim flow} shows the comparison of the three protocols as a function of number of sources in the network. For all cases, sources are different routers. For each flow, one specific router is selected as the destination with probability 0.5 and a random destination is selected with probability 0.5. 

The PDR decreases and the average end-to-end delays increase for all three protools as the number of sources increases. These results can be explained from the additional congestion created in the channel as a result of having  more data packets when more sources are added.  The results for  signaling overhead as a function of the number of sources  clearly show the benefits of RREQ aggregation in ADARA compared to AODV and OLSR. Although the signaling traffic in ADARA does increase as the number of sources increases, many of those sources share common destinations and this results in many RREQs being aggregated, which in turn results in much smaller overhead than with the other two protocols.

\subsection{Effect of Network Size}

Figure \ref{sim networkSize} shows the performance of the three protocols as a function of the number of routers. We considered different network sizes of 25, 50, 75, and 100 routers spread randomly in a area of $300m \times 1000m$, $300m \times 1500$, $300m \times 2000$, and $500m \times 2200$ with  15, 25, 40 and 50 flows respectively. Similar to the previous scenarios, a destination is a specific router with probability of 0.5 and it chosen randomly with probability of 0.5.

As we have stated, the scenarios were selected to stress all protocols, rather than to show likely operating points. For all three protocols, as the network size increases the 
PDR  drops, end-to-end delays increase, and signaling overhead increases.
This is unavoidable, given that OLSR must send more link states, and AODV and ADARA must send more route requests as the network size increases. However, it is clear that ADARA is more scalable than OLSR and AODV, and is far more efficient than AODV.

\section{Conclusions}

We introduced   route-request aggregation as an effective mechanism to significantly reduce the signaling overhead incurred in route discovery, and  presented ADARA as an example of the basic approach. 

ADARA  uses destination-based sequence numbers to avoid routing-table loops like AODV does,  and uses RREQ aggregation and broadcast signaling packets to reduce signaling overhead.  We compared the performance of ADARA, AODV, and  OLSR and  analyzed the effect of mobility, number of flows, and network size on the performance of the protocols. The simulation   results show that, in terms of packet delivery ratio, ADARA performs much better than AODV and performs very close to OLSR in all cases. The signaling overhead incurred with ADARA is much smaller than the overhead in AODV and OLSR. Furthermore, the use of RREQ aggregation and broadcast signaling packets  in ADARA leads to fewer packets contenting for the channel and  results in lower end-to-end delays for  ADARA compared to AODV and OLSR. 

As we have stated, the basic approach of using RREQ aggregation can be applied to any on-demand routing protocol. Accordingly, our results offer a great opportunity to improve the performance of on-demand routing protocols being considered for standardization. Our results indicate that RREQ aggregation can make AODV, DSR, or other on-demand routing protocols, far more attractive compared to OLSR and other proactive routing protocols.

 Our description of ADARA  assumed single-path routing. However, multi-path routing \cite{mosko, roam} can be easily supported as well.  Furthermore, as we we have stated, RREQ aggregation  can be used together with other techniques that have been proposed to improve the performance of on-demand routing in MANETs.  
 
 The next steps for our work on unicast routing include the definition and analysis of multi-path routing based on ADARA, the use of loop-avoidance techniques other than destination-based sequence numbers in the context of route-request aggregation, the use of geographical coordinates as in \cite{lar, wang-13}, and the use of clustering techniques. In addition, it is clear that the use of route-request aggregation can be applied to improve the performance of on-demand multicast routing  \cite{azzedine, prime}.

\section{Acknowledgments}

This work was supported in part by the Baskin Chair of Computer Engineering at UCSC.

%\newpage
% \vspace{-0.05in}
 {\fontsize{8}{8}\selectfont

  }

\end{document}